\documentstyle[a4,11pt,amssymb]{article}

\newcommand{\be}{\begin{equation}}
\newcommand{\ee}{\end{equation}}
\newcommand{\Real}{\mbox{Re}}
\newcommand{\Tr}{\mbox{Tr}}
\newcommand{\bea}{\begin{eqnarray}}
\newcommand{\eea}{\end{eqnarray}}

\newcommand{\nn}{\nonumber}

\newcommand{\link}{\begin{array}{l}\begin{picture}(85,125)
	\put ( 0, 0 ) 	{\circle{1}}
	\put ( 0,40 ) 	{\circle{1}}
	\put ( 0,80 ) 	{\circle{1}}

	\put (40, 0 ) 	{\circle{1}}
	\put (40,40 ) 	{\circle{1}}
	\put (40,80 ) 	{\circle{1}}

	\put (80, 0 ) 	{\circle{1}}
	\put (80,40 ) 	{\circle{1}}
	\put (80,80 ) 	{\circle{1}}

	\put (120, 0) 	{\circle{1}}
	\put (120,40) 	{\circle{1}}
	\put (120,80) 	{\circle{1}}
	
	\put (36,28) {$x$}
	\put (70,28) {$x+a\mu$}

	\put (40,40) {\vector(1,0){23}}
	\put (40,40) {\line(1,0){40}}

\end{picture}\end{array}}
\newcommand{\plaquette}{\begin{array}{l}\begin{picture}(22,25)

	\put (0,0) {\vector(1,0){13}}
	\put (0,0) {\line(1,0){20}}
	\put (20,0) {\vector(0,1){13}}
	\put (20,0) {\line(0,1){20}}
	\put (20,20) {\vector(-1,0){13}}
	\put (20,20) {\line(-1,0){20}}
	\put (0,20) {\vector(0,-1){13}}
	\put (0,20) {\line(0,-1){20}}
	\put (8,-9) {\em i}
	\put (23,8) {\em j}
	
\end{picture}\end{array}}
\newcommand{\rectangleA}{\begin{array}{l}\begin{picture}(42,25)

	\put (0,0) {\line(1,0){20}}
	\put (0,0) {\vector(1,0){13}}
	\put (20,0) {\line(1,0){20}}
	\put (20,0) {\vector(1,0){13}}	
	\put (40,0) {\vector(0,1){13}}
	\put (40,0) {\line(0,1){20}}
	\put (40,20) {\vector(-1,0){13}}
	\put (40,20) {\line(-1,0){20}}
	\put (20,20) {\vector(-1,0){13}}
	\put (20,20) {\line(-1,0){20}}
	\put (0,20) {\vector(0,-1){13}}
	\put (0,20) {\line(0,-1){20}}
	\put (8,-9) {\em i}
	\put (43,8) {\em j}	

\end{picture}\end{array}}

\begin{document}
\vspace{6cm}
\begin{center}
{\Large\bf Direct Improvement of Hamiltonian Lattice Gauge Theory}\\
\vspace{1cm}
{\large Jesse Carlsson\footnote{e-mail:
j.carlsson@physics.unimelb.edu.au}
 and Bruce H. J. 
 McKellar\footnote{e-mail: b.mckellar@physics.unimelb.edu.au}}\\
{\it School of Physics}\\
{\it The University of Melbourne}\\
\vspace{.3cm}
May 14, 2001 \\
\vspace{1.5cm}
{\bf Abstract}\\
\vspace{.5cm}
\begin{minipage}{13cm}{
We demonstrate that a direct approach to improving Hamiltonian
lattice gauge theory is possible. 
Our approach is to correct errors in the Kogut-Susskind Hamiltonian by 
incorporating additional gauge invariant terms. 
The coefficients of these terms are chosen
so that the order $a^2$ classical errors vanish. 
We conclude with a brief discussion of tadpole improvement in
Hamiltonian lattice gauge theory.}
\end{minipage}\\
\end{center}
\vspace{1.5cm}

\section{Introduction}

The idea of using the lattice as an ultra-violet regulator for quantum
chromodynamics (QCD) 
was proposed by Wilson in 1974 in his action formulation of
lattice gauge theory~\cite{Wilson:1974sk}. 
Soon after, Kogut and Susskind formulated the corresponding Hamiltonian
version of lattice gauge theory~\cite{Kogut:1975ag}.  
Both approaches were developed by demanding the correct continuum
limit be obtained in the limit of vanishing lattice spacing $a$. 
Creutz showed that the
Kogut-Susskind Hamiltonian could be derived from the Wilson action
using the transfer matrix method~\cite{Creutz:1983mg}. 
Later, Kogut demonstrated that the same
could be done by taking the continuous time limit of the Wilson action
and performing a canonical Legendre
transformation~\cite{Kogut:1983ds}.\\ 
  
To date, the majority of work in lattice QCD has been
performed in the action formulation. An advantage of this
approach is that it readily lends itself to Monte Carlo techniques.
Working in the Hamiltonian approach brings a different
intuition to the problem and serves as a check of universality. An
advantage of Hamiltonian lattice gauge theory is in the applicability of 
 techniques from
many body physics~\cite{McKellar:2000zk}. Also, it appears that in finite
density QCD, a Hamiltonian approach is favourable due to the so-called
complex action problem which rules out the use of standard 
Monte Carlo techniques in the action formulation~\cite{Gregory:2000pm}.\\

Much work in the past decade has been devoted to improving 
lattice actions~\cite{Lepage:1996jw}. 
Initiated by Symanzik in 1983~\cite{Symanzik:1983dc}, 
the aim of the
improvement programme is to reduce the deviation
between lattice and continuum QCD. For pure gauge theory on a
lattice, the deviations between continuum and lattice theories start at
order $a^2$. The motivation for
improvement lies in the fact that the computational 
cost of a lattice QCD simulation
is proportional to $a^{-k}$, where $6<k<7$.
It is by far more efficient to build an improved theory than it is to
work on finer lattices.
The improvement programme has
allowed accurate calculations to be 
performed on relatively coarse lattices and brought the most
complicated calculations within the reach of today's most powerful
computers.\\

In contrast, the improvement of lattice
Hamiltonians has only recently begun. Perhaps the most extensive treatment to
date is due to Luo, Guo, Kr\"oger and Sch\"utter~\cite{Luo:1999dx} 
who discussed the improvement
of Hamiltonian lattice gauge theory for gluons.
In their study it was discovered that deriving an
improved Hamiltonian from a Symanzik improved action, whether by transfer
matrix or canonical Legendre transformation, results in a kinetic Hamiltonian
with an infinite number of terms coupling lattice sites which are
arbitrarily far apart. To derive a local kinetic
Hamiltonian coupling only nearest neighbour lattice sites it was found
necessary to start with an improved action with an infinite number
of terms, coupling distant lattice sites.\\

With this technique the order $a^2$ errors are removed from the
Kogut-Susskind Hamiltonian. However,
generating Hamiltonians with further improvement would seem exceedingly
difficult. This is because one would need to start from a
L\"uscher-Weisz improved action with non-planar terms~\cite{Luscher:1985zq}.
For this reason we propose a move to the Symanzik approach, as applied 
to the Hamiltonian.
That is, to construct improved Hamiltonians directly by adding appropriate
gauge invariant terms and fixing their coefficients so that errors are
cancelled. \\

To date we have implemented Symanzik improvement to order $a^2$ in the
pure lattice gauge theory 
Hamiltonian (and to order $a^4$ in the kinetic part). 
We report those results here\footnote{The $O(a^{2})$ results 
were given in preliminary form in~\cite{McKellar:2000gp}}.\\

\section{Errors in Lattice Gauge Theory}\label{errors}

Before discussing Hamiltonian improvement we must first understand
how deviations between lattice gauge theory and its continuum
counterpart arise. The deviations can be separated
into two classes, classical and quantum errors, which will be described
in what follows.\\

\begin{figure}
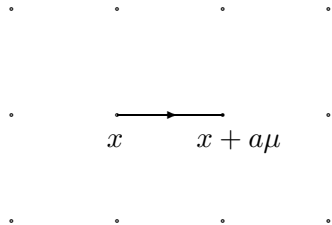

\bea
\link \nn
\eea
\caption{The diagrammatic representation of the link operator $U_\mu(x)$.}
\label{link}
\end{figure}

Rather than being
constructed from gluon fields, pure gauge theory on
the lattice is built from link operators 
\bea
U_\mu(x) = \exp \left\{ig\int_{\mbox{\tiny Link}} dx\cdot A\right\}.
\label{linkoperator} 
\eea
Here $g$ is the QCD coupling, 
$\mu=0,\ldots, 3$ is the Dirac index labelling the time-like and
space-like directions, $x$ labels a lattice site and $A$
is the gluon field. The integral runs along the link joining the
lattice sites $x$ and $x+a\mu$. This leads to the 
diagrammatic representation of the link operator shown in figure~\ref{link}. 
Lattice gauge theories are built from link operators rather than gluon
fields in order to maintain
manifest gauge invariance. \\

We define the $lattice$ gluon field $A^L_\mu(x')$ to be the average of the
continuum gluon field $A$ along the link joining $x$ and $x+a\mu$:
\bea
A^L_\mu(x') = \frac{1}{a} \int_{\mbox{\tiny Link}} dx\cdot A
\quad\Rightarrow \quad U_\mu(x) = e^{iga A^L_\mu(x')}, \label{latticegf}
\eea 
where $x'$ is a point near the points $x$ and $x+a\mu$.
On the lattice the gluon field is defined at only one
point along (or nearby) a link. This leads to interpolation errors in
the integral in eqn~\ref{latticegf}. For instance, by choosing to
evaluate the gluon field at the midpoint of the link,
the lattice and continuum fields are related by
\bea
A^L_\mu(x) = A_\mu(x) + \frac{a^2}{24}\partial_\mu^2 A_\mu(x) + 
\frac{a^4}{1920}\partial_\mu^4 A_\mu(x) + \dots
\eea
We see that the lattice gluon field reduces to its continuum
counterpart in the continuum limit ($a\rightarrow 0$), but that they 
differ by terms of order $a^{2}$, which may be called interpolation 
errors. \\

Having discussed classical errors we now move on to quantum errors in
lattice gauge theory. 
Quantum errors arise in two different contexts. Firstly, the
lattice acts as an ultraviolet regulator 
allowing the simulation
of only those 
states with momenta less than $\pi/a$. The absence of high
momentum states results in a deviation between lattice and
continuum theories. Secondly, non-physical
interactions arise due to the use of the link operator in constructing
the lattice theory. 
To demonstrate this we expand the link operator in powers of $g$, 
\be
 U_\mu(x) = 1 + iga A_\mu(x) -\frac{g^2a^2}{2!} A_\mu(x)  A_\mu(x) + \dots, 
\ee 
and note that the interaction of {\em any} number of gluons is
allowed. Naively, the unphysical interactions are suppressed by
powers of $a$. However, when contracted, products of pairs of gluon fields
produce ultraviolet divergences ($\propto 1/a^2$) which {\em exactly}
cancel the $a$ dependence of the expansion. These terms can be
uncomfortably large and result in what are known as tadpole errors.\\

In the last decade the improvement programme has led to a good
understanding of both classical and quantum errors in quark and
gluon actions
 (See reference~\cite{Lepage:1996jw} and references within). 
In contrast, only the lowest order
classical errors have been corrected in the Kogut-Susskind
Hamiltonian. Conjectures have been made about the structure of a
quantum improved Hamiltonian~\cite{Luo:1999dx}, 
but a perturbative study has not yet been
carried out.\\

We briefly describe how quantum improved Hamiltonian can be constructed 
in \S~\ref{tad}.

\section{Symanzik Improvement of the Lattice Hamiltonian} 

\subsection{Introduction}

In this section we derive an improved Hamiltonian {\em directly} using
the Symanzik approach of adding irrelevant terms and fixing their
coefficients in order to cancel errors. As a first step we aim to
correct the classical order $a^2$ errors
arising in the lattice Hamiltonian for pure SU($N$) gauge theory.\\

The Kogut-Susskind Hamiltonian for pure SU($N$) gauge theory on the
lattice is given by
\bea
H^{(0)} = K^{(0)} + V^{(0)},\label{ks} 
\eea
where the kinetic and potential terms are given respectively by,
\bea
K^{(0)} &=& \frac{a^3}{2} \sum_{x,i} \Tr \left\{E^L_i(x) E^L_i(x)
\right\}\\ 
V^{(0)} &=& \frac{2N}{ag^2}\sum_{x, i<j} P_{ij}(x). 
\eea
Here $E^L$ is the lattice chromo-electric field, $N$ is the dimension
of the gauge group, and $P_{ij}(x)$ is the
plaquette operator in the $(i,j)$ plane,
\bea
	P_{ij}(x) = 1- \frac{1}{N}\Real\Tr\left\{\!\!\plaquette\right\}.
\eea
We discuss the improvement of the kinetic and potential terms 
separately in what follows.\\

\subsection{Improving the Potential Term}

To improve the potential part of the Kogut-Susskind Hamiltonian, we follow
the process of improving the Wilson action. 
Our first step is to expand the potential term in
powers of $a$. This is done using Stokes' theorem to expand the
plaquette operator,
\bea
P_{ij}(x) &=& 1- \frac{1}{N}\Real\Tr\exp\left\{ig\oint_{\Box} A\cdot
dx\right\}\nn\\ 
&\approx &  1- \frac{1}{N}\Real\Tr\exp 
\left\{ig\int_{-a/2}^{a/2}\int_{a/2}^{a/2} dx'_i dx'_j 
\left[D_iA_j(x+x')- D_jA_i(x+x') \right]\right\}\nn\\
&\approx & \frac{g^2a^4}{2N}\Tr \left\{F_{ij}(x)F_{ij}(x)\right\} +
\frac{g^2a^6}{12N}\Tr \left\{F_{ij}(x)(D_i^2+D_j^2)F_{ij}(x)\right\}
+\dots \label{pij}
\eea
Here $D_i = \partial_i - ig A_i$ is the gauge covariant derivative, and 
$F_{ij}$ are the spatial component of the gluon field tensor.
The potential term then becomes:
\bea
V^{(0)} &=& \frac{2N}{ag^2}
\sum_{x, i<j} P_{ij}(x) \nn\\
&=& a^3\sum_{x, i<j} \left\{\Tr F_{ij}(x)F_{ij}(x) +
\frac{a^2}{6}\Tr F_{ij}(x)(D_i^2+D_j^2)F_{ij}(x) + \dots\right\}\nn\\
&\approx &  \frac{1}{2}\int d^3x \sum_{ij}\Tr \left\{F_{ij}(x)F_{ij}(x)
+\frac{a^2}{12} F_{ij}(x)(D_i^2+D_j^2)F_{ij}(x) + \dots\right\}.
\eea
We see that the correct continuum limit is restored in the limit as
$a\rightarrow 0$.
To cancel the order $a^2$ error we introduce rectangular loops into
the potential term. These are the next most complicated gauge
invariant constructions on the lattice after plaquettes. We write the improved
potential term as the linear combination
\bea
V^{(1)} = \frac{2N}{ag^2}\sum_{x,i<j} \left[XP_{ij}(x) +
\frac{Y}{2}\left(R_{ij}(x)+R_{ji}(x)\right)\right] ,\label{v1}
\eea
where $R_{ij}(x)$ is the rectangle operator in the $(i,j)$ plane with 
the long side in the $i$ direction, given by:
\bea
R_{ij}(x) = 1- \frac{1}{N} \Real\Tr\left\{\!\! \rectangleA\right\}.
\eea
The constants $X$ and $Y$ are to be fixed so that the order $a^2$
errors vanish. \\

The rectangle operator can be
expanded in powers of $a$ using Stokes' theorem:
\bea
R_{ij}(x) &=& \frac{4g^2a^4}{2N}\Tr \left\{F_{ij}(x)F_{ij}(x)\right\} +
\frac{4g^2a^6}{24N}\Tr \left\{
F_{ij}(x)(4D_i^2+D_j^2)F_{ij}(x)\right\}+\dots \label{rij}
\eea
Substituting eqns~\ref{pij} and~\ref{rij} into eqn~\ref{v1} we see
that cancelling the order $a^2$ error in $V^{(1)}$ 
requires $X=5/3$ and $Y=-1/6$. This leads to the improved potential term
\bea
V^{(1)} = \frac{2N}{ag^2}\sum_{x,i<j} \left[\frac{5}{3} P_{ij}(x) -
\frac{1}{12}\left(R_{ij}(x)+R_{ji}(x)\right)\right] . \label{vimp1}
\eea

In principle, the next lowest order classical errors
 could be corrected by
including additional, more complicated terms in the potential
term. This has not been done because many additional diagrams are
required to cancel the large number of order $a^4$ error terms. 
Since these errors are swamped by order $a^2 g^2$ quantum
errors, addressing quantum corrections in the
Hamiltonian approach would seem to be of more immediate importance.

\subsection{Improving the Kinetic Term}
 
Constructing a kinetic Hamiltonian with a finite number of terms has
proven to be a nontrivial exercise. Luo, Guo, Kr\"oger and Sch\"utter
demonstrated an interesting trade off when using either the transfer
matrix or Legendre transformation methods to derive an improved
Hamiltonian~\cite{Luo:1999dx}. Both techniques
require the starting point to be an improved action. 
When one starts from an improved action incorporating rectangular terms 
the resulting Hamiltonian has infinitely many terms and couples links 
which are arbitrarily far apart. To produce a Hamiltonian which
couples only nearest neighbour links, it was found necessary to start
from a carefully constructed highly non-local improved action.\\

Here we demonstrate an alternative approach, similar in nature to the
Symanzik improvement of the Wilson action. 
One only needs
to include additional gauge invariant terms with appropriate continuum
behaviour in the kinetic Hamiltonian.
The coefficients of the additional terms are chosen 
so that the order $a^2$ errors vanish. \\

An important step in understanding the errors that arise in the
kinetic Kogut-Susskind Hamiltonian involves making the distinction 
between {\em lattice} and {\em continuum} fields. In \S~\ref{errors}
 we defined the lattice gluon field on a link to be the average of the
continuum gluon field along the link:
\bea
A^L_\mu(x) = \frac{1}{a} \int_{\mbox{\tiny Link}} dx\cdot A = A_\mu(x)
+ \frac{a^2}{24}\partial_\mu^2 A_\mu(x) +  
\frac{a^4}{1920}\partial_\mu^4 A_\mu(x) + \dots 
\eea
From this we can build the following sequence of approximations to the
lattice gluon field: 
\bea
A^{(0)}_i(x) &=& A_i(x) \nn\\
A^{(1)}_i(x) &=& A_i(x) + \frac{1}{24}a^2 \partial^2_i A_i(x) 
\label{gapprox}\\
A^{(2)}_i(x) &=& A_i(x) + \frac{1}{24}a^2 \partial^2_i A_i(x) +
\frac{1}{1920}a^4 \partial^4_i A_i(x)\nn\\
	     &\vdots& \nn
\eea
Perhaps the most important property of the electric field is that it
generate group transformations. Mathematically, this translates
to the electric and gluon fields satisfying the commutation relations,
\bea
[E^\alpha_i(x), A^\beta_j(y)] = -\frac{i}{a^3} \delta_{xy}
\delta_{ij}\delta_{\alpha\beta}.
\eea
It is desirable for this hold on the lattice for any degree of
approximation. Let us consider what happens to these commutation
relations on the lattice for the approximation labelled by the
superscript (1) in eqns~\ref{gapprox}:
\bea
[E^{(1)\alpha}_i(x), A^{(1)\beta}_j(y)] =[E^{(1)\alpha}_i(x),
A^{\beta}_j(y)+  \frac{a^2}{24}\partial_j^2 A^\beta_j(y)] . 
\eea
We observe that if
the lattice electric field is taken to be the continuum electric
field, order $a^2$ errors arise in the commutation relations. To
cancel this error we set
\bea
E^{(1)\alpha}_i(x) = E^\alpha_i(x) - \frac{a^2}{24}\partial_i^2
E^\alpha_i(x). \label{e1}
\eea
We can take this to order $a^4$ by setting
\bea
E^{(2)\alpha}_i(x) = E^\alpha_i(x) - \frac{a^2}{24}\partial_i^2
E^\alpha_i(x) 
+ \frac{7a^4}{5760}\partial_i^4 E^\alpha_i(x). \label{e2}
\eea
In this way a sequence of approximations to the {\em lattice}
electric field $E^L$ can be constructed.\\

Making use of these approximations we can analyse the   
classical errors arising in the kinetic Hamiltonian. 
To cancel these errors we take the approach of adding new terms and
fixing their coefficients in order to cancel the order $a^2$ error. 
We have a great deal of freedom in choosing additional terms. They are
restricted only by gauge invariance and the need for an appropriate continuum 
limit.\\

To understand the construction of gauge invariant kinetic terms it is
important to recall that the electric field and link operator
transform as follows under a local gauge transformation $\Lambda(x)$:  
\bea
E_i(x) &\rightarrow & \Lambda(x) E_i(x) \Lambda^\dagger(x) \\
U_i(x) &\rightarrow & \Lambda(x) U_i(x) \Lambda^\dagger(x+ai).
\eea
Consequently, the next most
complicated gauge invariant term we can construct (after $\Tr E^LE^L$)
couples nearest neighbour electric fields:
\bea
\Tr \left\{E_i(x) U_i(x) E_i(x+ai) U_i^\dagger(x)\right\}.
\eea 
More complicated gauge invariant terms are easily constructed. One
only needs to couple electric fields on different links anywhere
around a closed loop. Consequently, generating Hamiltonians with
higher degrees of improvement
would be seem to be more readily achieved within this approach. \\

Incorporating nearest neighbour interactions leads to the 
simplest improved kinetic Hamiltonian: 
\bea
K^{(1)} = \frac{a^3}{2} \sum_{x,i} \Tr\left\{ X E^L_i(x)E^L_i(x) + Y
E^L_i(x) U_i(x) E^L_i(x+ai) U_i^\dagger(x)\right\}.
\eea
We fix the coefficients $X$ and $Y$ to cancel the order $a^2$
error. To do this we expand the second term in a Taylor series in $a$ 
. Ignoring ${\cal O}(g^2a^2)$
errors, we then substitute $E^L \approx E^{(1)}$ from eqn~\ref{e1}.
To cancel the order $a^2$ error we must set 
 $X = 5/6$ and $Y=1/6$. This results in the order $a^2$
improved kinetic Hamiltonian
\bea
K^{(1)} = \frac{a^3}{2} \sum_{x,i} \Tr\left\{ \frac{5}{6}
E^{L}_i(x)E^{L}_i(x) + \frac{1}{6}
E^{L}_i(x) U_i(x) E^{L}_i(x+ai) U_i^\dagger(x)\right\}. \label{kimp1}
\eea
This is the result of Luo, Guo, Kr\"oger and Sch\"utter~\cite{Luo:1999dx}.
We can take the this to order $a^4$ by including next nearest neighbour
interactions. A similar calculation using $E^L\approx E^{(2)}$ from
eqn~\ref{e2} gives 
\bea
K^{(2)} &=& \frac{a^3}{2}\sum_{x,i}\Tr\left\{ \frac{97}{120}E^L_i(x)E^L_i(x) 
+ \frac{1}{5}E^L_i(x)U_i(x)E^L_i(x+ai)U^\dagger_i(x) \right. \nn\\
&& \left. \hspace{1cm}- \frac{1}{120}
E^L_i(x)U_i(x)U_i(x+ai)E^L_i(x+2ai)U^\dagger_i(x+ai)U^\dagger_i(x)\right\}. 
\label{kimp2}
\eea

\section{Tadpole Improvement} \label{tad}

Tadpole improvement, developed by Lepage and Mackenzie~\cite{Lepage:1993xa},
is an important step in removing errors from
lattice gauge theory. It is necessary for close agreement between 
lattice perturbation theory and Monte Carlo calculations on coarse
lattices. \\

In the action formulation tadpole improvement is handled by dividing
all link operators by the mean link $u_0$. In the Hamiltonian approach
two conflicting implementations have been suggested. The earliest
 starts from a tadpole improved action and carries factors of $u_0$ into the
Hamiltonian~\cite{Luo:1999dx}. More recently it was suggested that no tadpole
improvement was necessary in the kinetic term of the improved
Hamiltonian~\cite{Fang:2000vm}. Here we present our own views on the correct
implementation.\\
 
In the Hamiltonian approach
the question of whether the electric field should be scaled
arises. This question is easily answered by considering the
commutation relations between the link operator and electric field:
\be
[E^\alpha_i(x), U_j(y)] = \frac{g}{a^2}
\delta_{ij}\delta_{xy}\lambda^\alpha U_i(x).
\ee 
We see that if we divide all link operators by $u_0$ we have
\be
[E^\alpha_i(x), \frac{1}{u_0}U_j(y)] =
\frac{g}{a^2}\delta_{ij}\delta_{xy}\lambda^\alpha \frac{1}{u_0}U_i(x). 
\ee
We observe that the electric field cannot be rescaled and still maintain the
correct commutation relations. Thus under tadpole improvement the
electric field cannot change. We must, however, divide the second of
the kinetic terms by a factor of $u_0^2$. Tadpoles arise in this term
because the electric and gluon fields do not commute.\\

Including tadpole improvement in eqns~\ref{vimp1} and~\ref{kimp1} 
leads to the order $a^2$ tadpole improved Hamiltonian: 
\bea
H^{(1)} &=& K^{(1)} +V^{(1)}\nn\\
 &=& \frac{a^3}{2} \sum_{x,i} \Tr\left\{\frac{5}{6}
E^{L}_i(x)E^{L}_i(x) + \frac{1}{6u_0^2}
E^{L}_i(x) U_i(x) E^{L}_i(x+ai) U_i^\dagger(x)\right\} \nn\\
&&-\frac{2N}{ag^2}\sum_{x,i<j} \left[\frac{5}{3u_0^4} P_{ij}(x) -
\frac{1}{12u_0^6}\left(R_{ij}(x)+R_{ji}(x)\right)\right].
\eea

\section{\bf Conclusion}

We have demonstrated that direct improvement of the Kogut-Susskind
Hamiltonian by demanding the correct continuum limit is possible.
The advantage of our direct approach is that it is easily extended
to more complicated Hamiltonians.
One simply needs to construct suitable gauge invariant terms to add to
the kinetic Hamiltonian and fix the coefficients so that higher order
errors are cancelled.\\

Our next step is to perform variational and coupled cluster
SU(3) calculations to determine precisely the level of improvement
achieved by the
improved Hamiltonians. Other groups have made progress
with these calculations for U(1)~\cite{Fang:2000vm} and
SU(2)~\cite{Li:2000bg} with promising results.\\

In the near future we intend to extend the
direct approach to the cancellation of quantum errors which have not
yet been examined in Hamiltonian lattice gauge theory.\\

\end{document}